\documentclass[fleqn,12pt,twoside]{article}
 \usepackage{espcrc1}
 \readRCS
$Id: espcrc1.tex,v 1.2 2004/02/24 11:22:11 spepping Exp $
\usepackage{graphicx}
\usepackage{amssymb}

\newcommand{\AmS}{{\protect\the\textfont2
  A\kern-.1667em\lower.5ex\hbox{M}\kern-.125emS}}
\newcommand{\efm}[1]{\newcount\efmpow
\efmpow=#1
\multiply\efmpow by 2
\ensuremath{\mathrm{e}^2\mathrm{fm}^{\number\efmpow}}}
\newcommand{\nuc}[2]{$^{#1}${#2}}
\newcommand{\Ne}[1]{\nuc{#1}{Ne}}
\newcommand{\Pb}{\nuc{nat}{Pb}}
\newcommand{\Al}{\nuc{nat}{Al}}
\newcommand{\Ar}{\nuc{40}{Ar}}

\newcommand{\gcm}{mg/cm$^2$}
\newcommand{\x}{$\times$}
\newcommand{\shift}{\ensuremath{\hookrightarrow}}

\newcommand{\ipno}{\address[ipno]{Institut de Physique Nucl\'eaire, IN$_2$P$_3$-CNRS,
    F-91406 Orsay, France}}
\newcommand{\rikkyo}{\address[rikkyo]{Department of Physics, Rikkyo University, 3-34-1
    Nishi-Ikebukuro, Toshima, Tokyo 171-8501, Japan}}
\newcommand{\riken}{\address[riken]{RIKEN (The Institute of Physical and Chemical
    Research), 2-1 Hirosawa, Wako, Saitama 351-0198, Japan}}
\newcommand{\atomki}{\address[atomki]{Institute of Nuclear Research of the Hungarian
    Academy of Sciences, PO Box 51, H-4001 Debrecen, Hungary}}
\newcommand{\titech}{\address[titech]{Department of Physics, Tokyo Institute of Technology,
    Tokyo 152-8551, Japan}}
\newcommand{\cns}{\address[cns]{Centre for Nuclear Study, University of Tokyo, RIKEN
 Campus, 2-1 Hirosawa, Wako, Saitama 351-0198, Japan}}
\newcommand{\lbl}{present address: Lawrence Berkeley National Laboratory,
  CA 94720 Berkeley, USA}

\begin{document}

\title{Search for low lying dipole strength  in the neutron rich
nucleus $^{26}$Ne}

 \author{
 J.~Gibelin\ipno\rikkyo\thanks{\lbl}
 D.~Beaumel\addressmark[ipno],
 T.~Motobayashi\riken,
 N.~Aoi\addressmark[riken],
 H.~Baba\addressmark[riken],
 Y.~Blumenfeld\addressmark[ipno],
 Z.~Elekes\atomki,
 S.~Fortier\addressmark[ipno],
 N.~Frascaria\addressmark[ipno],
 N.~Fukuda\addressmark[riken],
 T.~Gomi\addressmark[riken],
 K.~Ishikawa\titech,
 Y.~Kondo\addressmark[titech],
 T.~Kubo\addressmark[riken],
 V.~Lima\addressmark[ipno],
 T.~Nakamura\addressmark[titech],
 A.~Saito\cns,
 Y.~Satou\addressmark[titech],
 E.~Takeshita\addressmark[rikkyo],
 S.~Takeuchi\addressmark[riken],
 T.~Teranishi\addressmark[cns],
 Y.~Togano\addressmark[rikkyo],
 A.~M.~Vinodkumar\addressmark[titech],
 Y.~Yanagisawa\addressmark[riken],
 K.~Yoshida\addressmark[riken]
}

% typeset front matter
\maketitle

\begin{abstract}
 Coulomb excitation of the exotic neutron-rich nucleus \Ne{26} on a
\Pb\ target was measured at 58~A.MeV in order to search for low-lying
E1 strength above the neutron emission threshold. Data were also taken
on an \Al\ target to estimate the nuclear contribution. The
radioactive beam was produced by fragmentation of a 95~A.MeV \Ar\ beam
delivered by the RIKEN Research Facility. The set-up included a NaI
gamma-ray array, a charged fragment hodoscope and a neutron
wall. Using the invariant mass method in the \Ne{25}+n channel, we
observe a sizable amount of E1 strength between 6 and 10~MeV. The
reconstructed \Ne{26} angular distribution confirms its E1 nature. A
reduced dipole transition probability of $B(E1)~=~0.49\pm0.16$~\efm{1}
is deduced. For the first time, the decay pattern of low-lying
strength in a neutron-rich nucleus is obtained. The results are
discussed in terms of a pygmy resonance centered around 9~MeV.
\end{abstract}

%\begin{keyword}
%Coulomb excitation \sep Pygmy resonance \sep Radioactive beam \PACS
%25.70.De \sep 25.60.-t
%\end{keyword}

\section{Introduction}

Giant Resonances are a general feature of nuclei and their properties
give us a handle on the macroscopic and microscopic behavior of
nuclear matter. These modes have been extensively studied in stable
nuclei over the last 50~years and the recent inception of Radioactive
Ion Beam facilities opens the opportunity to extend these
investigations to exotic nuclei. Far from stability new modes are
expected to appear. In particular, the dipole response of neutron-rich
nuclei may exhibit strength at energies lower than the standard Giant
Dipole Resonance (GDR) often depicted as the oscillation of a deeply
bound core against a neutron skin giving rise to a so-called pygmy
resonance. Recent studies on Oxygen~\cite{leisten:oxygens:gdr} and
Tin~\cite{adrich:sn:pdr} isotopes have given the first experimental
indications for such an effect, while at the same time calling into
question its nature.

The theoretical approach is often based on mean field
calculations. Relativistic random phase approximation (RRPA) and
quasi-particle RRPA (QRRPA) calculations have been carried out by
{Cao} and {Ma} in \Ne{26}~\cite{ma:private} which predict that below
10~MeV excitation energy, almost 5\% of the Thomas-Reiche-Kuhn (TRK)
energy weighted sum rule is exhausted by strength centered around 8
MeV. This region of energy is located between the one neutron and the
two neutron emission thresholds. Thus, in order to investigate this
prediction, we performed Coulomb excitation of \Ne{26} at intermediate
energies on a lead target and used the invariant mass method to
reconstruct the B(E1) strength from the \Ne{26}$\to$\Ne{25}$^*+n$
reaction.

\section{Experimental details}

The experiment was performed at the RIKEN Accelerator Research
Facility.  A secondary \Ne{26} beam was produced
through fragmentation of a 95~A.MeV \Ar\ primary beam on a
2-mm-thick \nuc{9}{Be} target. The \Ne{26} was separated by the RIKEN
Projectile Fragment Separator (RIPS)~\cite{kubo:rips}. Particle
identification was unambiguously performed by means of the
time-of-flight (TOF) and the purity was 80\%. The \Ne{26} beam of
intensity $\sim 5.10^3$~pps and incident energy 58~A.MeV, was tracked
with two parallel-plate avalanche counters providing incident angle
and hit position onto the target. It then impinged alternatively on a
230~\gcm~\Pb\ and a 130~\gcm~\Al\ target.  Data obtained with aluminum
target are used in the following to estimate the contribution of
nuclear excitation to the data.

The outgoing charged fragments were measured using a set of telescopes
placed at 1.2~m downstream of the target. They consisted of two layers
(X and Y) of 500~$\mu$m silicon strip detectors (SSD) with 5~mm strips
which provided an total energy resolution of 1.5~MeV (FWHM). The last
layer used 3-mm-thick Si(Li) from the charged-particle detector
MUST~\cite{yorick:must}, provided a 9~MeV (FWHM) resolution on the
remaining energy (E). Unambiguous mass and charge identification of
all projectile like fragments was obtained using the E-$\Delta$E
method.

 In-beam gamma rays were detected using a 4$\pi$-gamma-detector,
DALI2~\cite{takesato:dali2}, which consists of 152 NaI detectors
placed around the target.  For 1.3~MeV gamma-rays, its measured
efficiency is approximately 15\% with an energy resolution of 7\%
(FWHM). The Doppler corrected gamma energy distribution obtained in
coincidence with the \Ne{25} isotope allows us to identify the gamma
decay from the adopted 1702.7(7)~keV and 3316.4(11)~keV excited
states. In addition we observed a $2063\pm67$~keV gamma-ray which
we related to the $2030\pm50$~keV excited state, only seen up to now
by transfer reaction~\cite{wilcox:25ne}.

 The hodoscope for neutron detection was an array of 4 layers of 29
plastic rods each, placed 3.5~m downstream of the target. Each layer
was composed of 13 [2.1~m\x6\x6~cm$^2$] and 16 [1.1~m\x6\x6~cm$^2$]
rods, arranged in a shape of a cross. Its total intrinsic efficiency
for the detection of one 60~MeV~neutron was calculated to be $\sim
25$\%.  Finally 29 thin plastic rods covered the front face of the
wall in order to veto charged particles as well as to provide an
active beam stopper. The neutron position is determined with an error
of $\pm$3~cm and the energy, from TOF information, with a 2.5~MeV
(FWHM) resolution for the neutrons of interest.

\section{Results}

We have performed a simulation of the experimental setup using the
Geant~3 package~\cite{brun:geant}. In order to check the reliability
of the analysis, we built the elastic scattering angular distribution
of \Ne{26} on \Pb\ at 55~A.MeV. It is in good agreement with a DWBA
calculation based on a \Ne{20}+\nuc{208}{Pb} at 40~A.MeV optical
potential from~\cite{beaumel:ne20}. We empirically generated optical
potential parameters for the \Ne{26}+\nuc{27}{Al} reaction and tested
them by comparing with the experimental elastic scattering.

Using the invariant mass method, the excitation energy of an unbound
state in the nucleus $^\mathrm{A}\mathrm{X}$ decaying to a state in
$^\mathrm{A-1}\mathrm{X}$ can be expressed by:
$E^* = E_\mathrm{rel} + S_n + {\textstyle\sum_i}\,E_{\gamma_{i}}$,
where $E_\mathrm{rel}$ is the relative energy between neutron and
$^\mathrm{A-1}\mathrm{X}$ and $S_n$ the one neutron emission
threshold. The gamma detection efficiency was not high enough to apply
an event-by event reconstruction technique. 
To extract the excitation energy spectrum, we hence used a method
based on adequate subtraction of relative energy spectra built for
these two kind of event.

The method can be illustrated in the schematic case where the daughter
nucleus $^\mathrm{A-1}\mathrm{X}$ has only one excited state below its
neutron threshold.  The excitation energy spectrum can be decomposed
in the sum of two contributions: the decay of $^\mathrm{A}\mathrm{X}$
to the ground-state (gs) of $^\mathrm{A-1}\mathrm{X}$ and the decay of
$^\mathrm{A}\mathrm{X}$ to the excited state (I) of
$^\mathrm{A-1}\mathrm{X}$.

The first contribution can be obtained by subtracting from the
inclusive relative energy spectrum $\left\lfloor
E_\mathrm{rel}(^\mathrm{A-1}\mathrm{X},n)\right\rceil $ the relative
spectrum of gamma-coincidence events $\left\lfloor
E_\mathrm{rel}(^\mathrm{A-1}\mathrm{X},n)|_\gamma\,\right\rceil $
divided by the gamma detection efficiency $\epsilon$, and shifted
``\shift'' by the one neutron emission threshold $S_n$ \emph{i.e.}:
$\displaystyle\left\lfloor E^*\right\rceil^{gs~} = \left\lfloor
E_\mathrm{rel}(^\mathrm{A-1}\mathrm{X},n)\right\rceil - \left\lfloor
E_\mathrm{rel}(^\mathrm{A-1}\mathrm{X},n)|_\gamma\,\right\rceil
/\epsilon \shift S_n $.

The second contribution is simply the relative energy spectrum of
gamma-coincidence events shifted by the gamma energy: $\left\lfloor
E^*\right\rceil^{(\mathrm{I})} = \left\lfloor
E_\mathrm{rel}(^\mathrm{A-1}\mathrm{X},n)|_\gamma\,\right\rceil
/\epsilon \shift S_n \shift E_\gamma$.  Finally, the excitation energy
spectrum for the \Ne{25}+n decay channel is given by: $\left\lfloor
E^*\right\rceil = \left\lfloor E^*\right\rceil^{gs~} +\left\lfloor
E^*\right\rceil^{(\mathrm{I})}$.  The method was tested by simulation
in this schematic case, and also in the realistic case taking into
account the experimental decay scheme of \Ne{25}.

\begin{figure}[ht]
\centering
\hspace{\stretch{1}}
\includegraphics[width=0.47\textwidth]{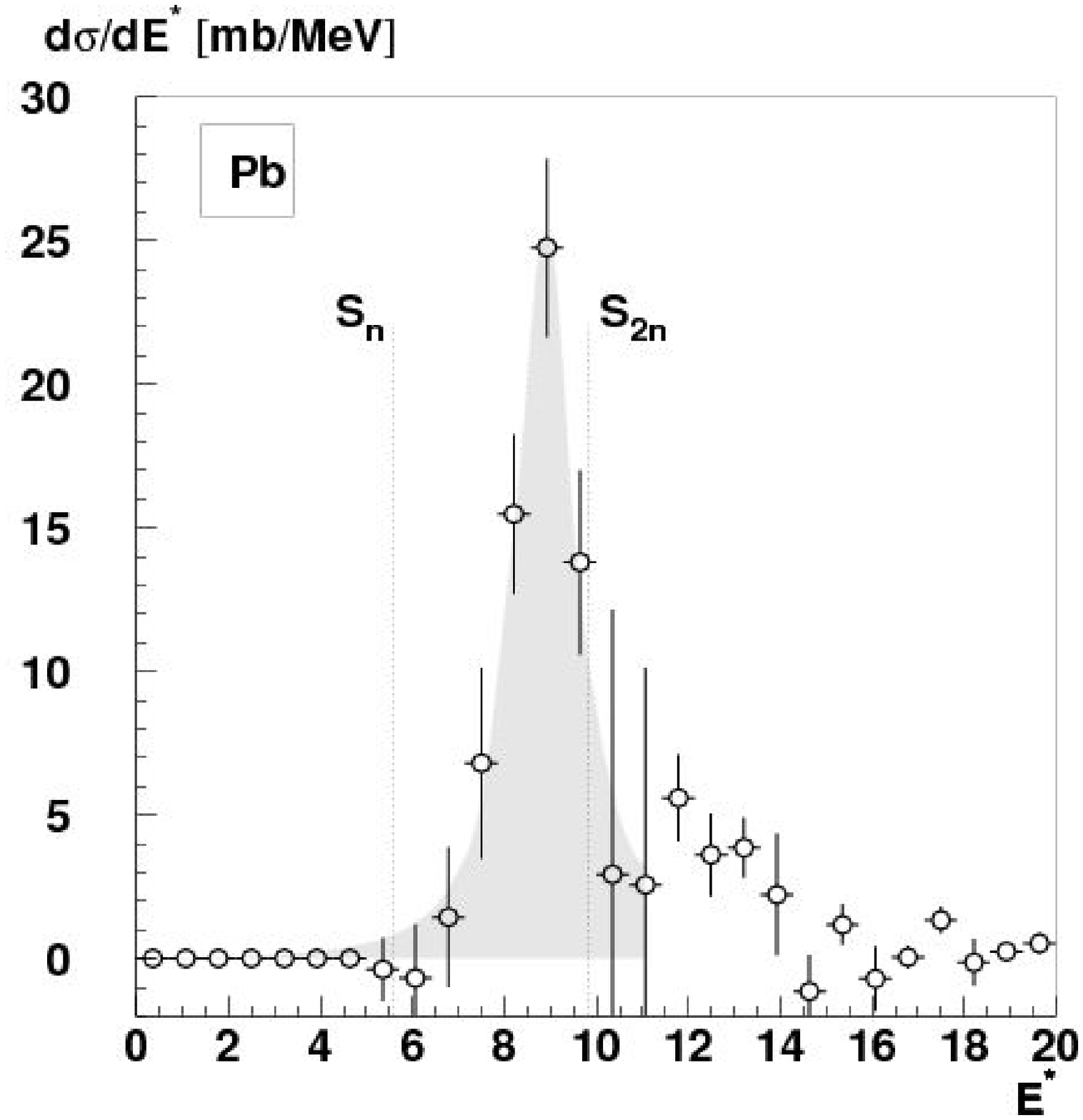}
\hspace{\stretch{1}}
\includegraphics[width=0.47\textwidth]{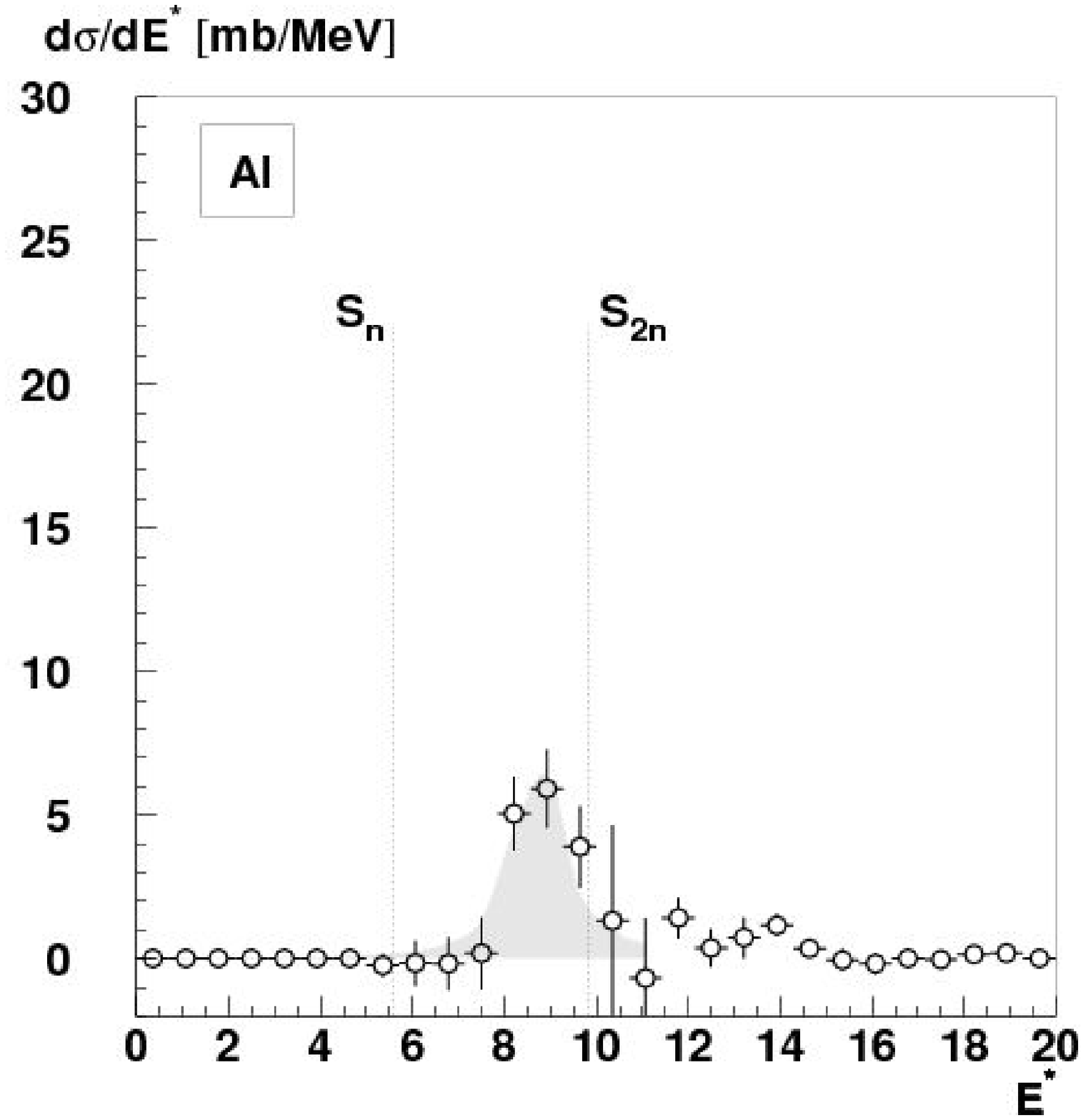}
\hspace{\stretch{1}}
\caption{\label{fig:invmas}{\bf Left:} Excitation energy distribution
in \Ne{26} in the \Pb\ target.  The shaded area is a tentative
Lorentzian fit.  {\bf Right:} Same as previous but for the \Al\
target.}
\end{figure}

The excitation energy spectra reconstructed for the \Ne{25}+n decay
channel obtained with the \Pb\ and \Al\ targets are represented on
Fig.~\ref{fig:invmas}. Note that above $S_{2n}=9.8$~MeV, the decay of
\Ne{26} is expected to occur mainly by 2 neutron emission.  Between 8
and 10 MeV, a sizable amount of cross-section is observed for both
targets. In intermediate energy inelastic scattering with a heavy
target such as \Pb, the Coulomb dominance of the E1 excitation is
well-known.  The contribution of possible E2 excitation to the
spectrum obtained with the lead target has --- in a first step ---
been determined using data taken with the aluminum target and the
coupled channels ECIS~97~\cite{ecis:code} code. Assuming simple
collective vibration mode with equal nuclear and Coulomb deformation
lengths, the E2 deformation parameters were extracted from the
measured cross-section with the \Al\ target ($\sigma_\mathrm{Al} =
9.1\pm2.3$~mb).  The $L=2$ cross section in lead was then calculated
using the deformation lengths extracted in the previous step. We
obtained $\sigma^{L=2}_\mathrm{Pb} = 17.9\pm4.3$~mb.  After
subtraction of the $L=2$ contribution, the resulting
$\sigma^{L=1}_\mathrm{Pb} = 48.5\pm4.8$~mb cross section corresponds,
using ECIS~97, to a Coulomb deformation parameter of $\beta_C =
0.087\pm0.05$ which led to a $B(E1)~=~0.55\pm0.05$~\efm{1} \emph{via}
the relation
$B(E1;0^+\to1^-)=\left(\frac{3}{4\pi}Z_p\,e\,R_C\beta^{L=1}_C\right)^2$
with $R_C$ the Coulomb radius. 
This value of reduced transition probability correspond to
$5.5\pm0.6$\% of the {Thomas-Reiche-Kuhn} energy weighted sum rule for
an excitation energy of 9~MeV.  We estimate the error due to the
choice of optical potential by performing the same analysis using now
parameters from the \nuc{40}{Ar}+\nuc{208}{Pb} reaction at
40~A.MeV~\cite{tiina:ar40}. We obtained $B(E1)=0.60\pm0.06$~\efm{2}
\emph{i.e.} $6.0\pm0.6$\% of the TRK, in good agreement with the
values previously extracted, which demonstrates that we are not
strongly sensitive to the choice of the optical potential for the
reaction on lead.

\begin{figure}[htb]
\begin{minipage}[b]{80mm}
\includegraphics[width=79mm]{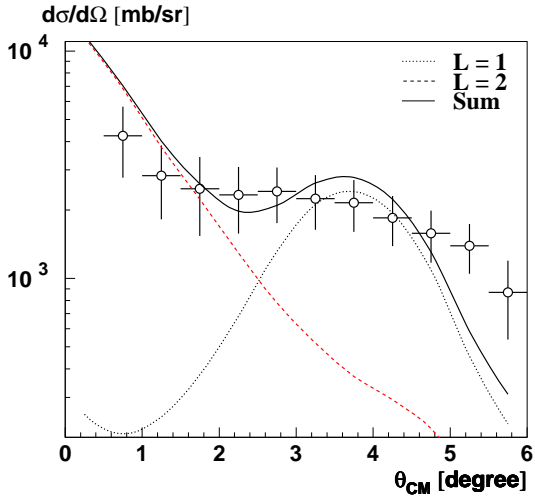}
\end{minipage} 
\raisebox{33mm}{\protect
\begin{minipage}[b]{70mm}
\caption{
\label{fig:pygmy:angdist:pb} Result (solid line) of the
multipole decomposition of the experimental angular distribution
(circle) of the \Ne{26} angular scattering onto lead target, by $L=1$
(dashed line) and $L=2$-type (dotted line) distributions.
}
\end{minipage}}
\end{figure}

Due to the high granularity and the good resolution of the present
setup, it is possible to reconstruct the scattering angular
distribution for \Ne{26} on the \Pb\ target which is represented in
Fig.~\ref{fig:pygmy:angdist:pb}.  Hence, we have used a second method
to extract the E1 excitation which relies on a multipole decomposition
analysis of this angular distribution. The $L=1$ and $L\geq2$ angular
distributions (dashed and dotted lines) were obtained from simulation
based on ECIS~97 angular distribution calculated for $E^*= 9$~MeV. The
data were fitted with a linear combination of the two
distributions. The results of the fit give us $B(E1)~=~0.49\pm0.16$
which corresponds to $4.9\pm1.6$\% of the TRK, again for an excitation
energy of 9~MeV. If we suppose now that the remaining part of the
contribution is due to $L=2$ excitation we can extracted a
$B(E2\uparrow)=49\pm8~\efm{2}$ for this structure.

The two results from the two different methods are in agreement but
since the multipole decomposition of the angular distribution relies
only on the data from lead target, the second value of $B(E1) =
0.49\pm0.16$~\efm{1} will be retained.  The results obtained from our
experiment concerning the E1 transition are compared in the following
to theoretical calculations.

%Note that the smaller
%error bar in the \emph{integrated} cross section method comes from the
%fact that the extrapolation factor from \Al\ to \Pb\ is theoretically
%calculated using ECIS~97 and its error not estimated.

\section{Comparison with theory}

In the introduction we already presented results from Cao and Ma
\cite{ma:private}. Using the relativistic QRPA framework and the
response function formalism they predicted a pygmy resonance centered
around 8.4~MeV and which exhausts $4.5\%$ of the TRK, which is close
to our experimental values.  Another calculation has been performed by
Khan~\emph{et~al.}  \cite{elias:priv}. It is based on effective SGII
Skyrme interactions and is performed in the spherical QRPA
framework. It predicts a redistribution of the strength at low energy
centered around $E^* = 11.7$~MeV exhausting $\sim 5\%$ of the TRK.
Two other preliminary calculations has been performed in the deformed
QRPA framework using Gogny forces~\cite{sophie:thisproc} and in the
deformed relativistic QRPA framework~\cite{ring:thisproc}.  Both
predict a redistribution of the strength at low energy, centered
around $10.7$~MeV and $7.5$~MeV respectively. The first calculation
also predicts that only $\sim 1\%$ of the TRK should be exhausted.
Preliminary shell-model calculations have also been performed by
Nowacki\emph{et~al.}~\cite{nowacki:private} who predict a state at
9.3~MeV exhausting $\sim5\%$ of the TRK.  All these theories agree on
the presence of a $1^-$ structure a low excitation energy, compatible
with our experimental result but they disagree on its nature: the
amount of strength differs as well as as well as its collective or
single particle nature.

\section{Decay of pygmy resonances in \Ne{26}}

\begin{table}[ht]
\caption[Branching ratios]{Experimental branching ratio compared to
statistical model for a given
\Ne{26} transition multipolarity to a given excited state in \Ne{25}
compared to experiment. \label{tab:stat:decay}}
\renewcommand{\tabcolsep}{1pc}
\renewcommand{\arraystretch}{1.1} % enlarge line spacing
\begin{tabular}{ccccccc}
\hline
\multicolumn{2}{c}{Final \Ne{25} state} &
\multicolumn{2}{c}{Experiment} &
\multicolumn{3}{c}{Statistical decay} \\
Label& $J^\Pi$      &  Pb           & Al            
& \ $L=1$\  & \ $L=2$\ & \ $L=3$\  \\
\hline
(gs) & $1/2^+$      &$5^{+17}_{-5}\%$ & $<10\%$      & 40\%  & 28\% & 22\%  \\
(I)  & $5/2^++3/2^+$&$66\%\pm15\%$  &$95^{+5}_{-15}\%$& 55\%  & 67\% & 75\%  \\
(II) & $3/2^-$      &$35\%\pm 9\%$  &$ 5^{+6}_{-5}\%$& 5\%   &  4\% &  3\%  \\
\hline
\end{tabular}
\end{table}

Our reconstruction method for the excitation energy allows us to
extract for the first time data on the decay of pygmy resonance of
neutron--rich nuclei.  We present the experimental branching ratios
towards the various states of \Ne{25} in Tab.~\ref{tab:stat:decay}.
The clear difference between the branching ratios obtained with lead
and aluminum targets proves that states of different nature have been
excited.  For comparison, we also performed a statistical decay
calculation for $L=1,2,3$ states using the CASCADE
code\cite{Puhlhofer77:cascade}, assuming spins and parities of \Ne{25}
states as listed in Tab.~\ref{tab:stat:decay}. It clearly shows that
the decay is not statistical, which is not surprising for a light
nucleus.  No predictions of the direct decay of pygmy states yet exist
from the previously mentioned microscopic models. Future comparisons
with our data should be a strong test for these models.

\section{Conclusion}

We performed the Coulomb excitation of \Ne{26} in order to measure its
low lying dipole strength below 10~MeV excitation energy, using the
invariant mass method.  We extract an E1 strength value of $B(E1) =
0.49\pm0.16$~\efm{1} between the one-- and the two--neutron emission
threshold as well as the corresponding decay pattern.  Our results are
compatible with various theoretical predictions. Future comparison of
the extracted decay pattern with theoretical models may allow us to
elucidate whether the strength is of single particle nature or rather
due to a collective pygmy resonance.

\end{document}